\documentclass{article}
\usepackage{graphicx}
\usepackage{subcaption}
\usepackage{amsmath,amsthm,amssymb,amsfonts,mathtools,bbold}
\usepackage{url}
\usepackage{cite}
\usepackage{hyperref}
\usepackage{enumitem}

\newcommand\numberthis{\addtocounter{equation}{1}\tag{\theequation}}

\begin{document}

\begin{titlepage}
{\ }
\vskip 1in

\begin{center}
{\LARGE{The horocycle regulator:  exact cutoff-independence in AdS/CFT}}
\vskip 0.5in {\Large Sristy Agrawal,$^a$ Oliver DeWolfe,$^b$ \\[0.3cm] Kenneth Higginbotham,$^b$  and Joshua Levin$^a$}
\vskip 0.2in 
{\it {}$^a$Department of Physics and JILA, University of Colorado/NIST, \\Boulder, CO, 80309, USA}
\vskip0.05in
{\it {}$^b$Department of Physics and Center for Theory of Quantum Matter, 390 UCB,  University of Colorado, \\ Boulder, CO 80309, USA}
\vskip0.2in
{\it Email:} sristy.agrawal@colorado.edu, oliver.dewolfe@colorado.edu,  \\kenneth.higginbotham@colorado.edu, jlevin1729@gmail.com
\end{center}

\vskip 0.7in

\begin{abstract}\noindent
% Now wondering if horocycles are more like motorcycles
While the entanglement entropy of a single subregion in quantum field theory is formally infinite and requires regularization, certain combinations of entropies are perfectly finite in the limit that the regulator is removed, the mutual information being a common example. For generic regulator schemes, such as a holographic calculation with a uniform radial cutoff, these quantities show non-trivial dependence on the regulator at finite values of the cutoff. We investigate a holographic regularization scheme defined in three-dimensional anti-de Sitter space constructed from \textit{horocycles}, curves in two-dimensional hyperbolic space perpendicular to all geodesics approaching a single point on the boundary,
that leads to finite information measures that are \textit{totally} cutoff-independent, even at finite values of the regulator. We describe a broad class of  such information measures, and describe how the field theory dual to the horocycle regulator is inherently non-local.
\end{abstract}

\end{titlepage}

\section{Introduction}

The AdS/CFT correspondence provides an exact ``holographic" duality between strongly coupled quantum field theories and theories of gravity in higher dimensional asymptotically anti-de Sitter (AdS) space, where the quantum field theory may be thought of as living on the boundary of the AdS space \cite{Maldacena:1997re,Gubser:1998bc,Witten:1998qj}. 
For example, for a particular state in a quantum field theory the entanglement entropy of a spatial region with its complement can be calculated in the gravity dual by finding the minimal area bulk surface sharing its boundary with the boundary of the region, and homologous to that region, according to the Ryu-Takayanagi (RT) formula \cite{ryu_holographic_2006},\footnote{The RT formula holds for time-independent states in the classical limit on the gravity side, corresponding to a limit of a large number of degrees of freedom for the dual field theory, both of which we will assume throughout; it can be generalized to general spacetimes \cite{hubeny_covariant_2007} and to include quantum corrections \cite{faulkner_quantum_2013,engelhardt_quantum_2015}, in which context the RT surface is called a quantum extremal surface. }
\begin{eqnarray}
\label{eq:RyuTakayanagi}
    S(A) = \frac{{\rm Area}(\Gamma_A)}{4G_N} \,,
\end{eqnarray}
where $G_N$ is Newton's constant, $A$ is the boundary spatial region, and $\Gamma_A$ is the minimizing bulk surface sharing a boundary with $A$, called the Ryu-Takayanagi surface. For convenience, let us use units where $4G_N = 1$, so that the entropy is simply the minimal surface area.

In practice, both sides of (\ref{eq:RyuTakayanagi}) are divergent, and must be regularized. On the gravity side, spacelike proper distances to the boundary where the field theory lives are infinite, and so the area of $\Gamma_A$ is formally infinite as well. Correspondingly, for the field theory there is an infinite amount of entanglement between any spatial region $A$ and its complement $\bar{A}$ localized along their mutual boundary, associated to the ultraviolet divergences of quantum field theory. Calculating the entropy requires introducing a regulator, and doing so on one side of the duality  implicitly introduces a cutoff in the dual description as well. In this paper we will explore the structure of these regulators, and how a certain regulator has interesting properties for calculating a class of observables derived from the entropy.

The example we will focus on is that of a two-dimensional conformal field theory (CFT), whose ground state is dual to pure three-dimensional anti-de Sitter space (AdS$_3$). For the CFT living on an infinite spatial line, the entanglement entropy of a region $A$ of length $L$ defined in the presence of a short-distance cutoff $\epsilon$ takes the form
\begin{equation}
    S(A) = 2 \ln {\frac{L}{\epsilon}} + {\cal O}(\epsilon) \,,
\end{equation}
which diverges when the cutoff is removed by taking the $\epsilon \to 0$ limit. However, certain combinations of these divergent entropies remain finite in the limit the cutoff $\epsilon$ is removed. A familiar example is the mutual information of two subregions $A$ and $B$,
\begin{equation}
    I(A:B) = S(A) + S(B) - S(AB)\,.
\end{equation}
So long as the subregions $A$ and $B$ do not share a common boundary, the divergent portions of $S(A)$, $S(B)$, and $S(AB)$ cancel in $I(A:B)$, leaving a finite result, along with cutoff-dependent corrections vanishing as the cutoff is removed. Such information measures may be called {\em finite}.

Although the mutual information is finite, the subleading corrections that arise with a generic regulator mean that the calculated value depends on the cutoff, with the physical result emerging as the cutoff is removed. The goal of this note is to show how with a particular holographic regulator for a two-dimensional  CFT, the {\em horocycle regulator}, these finite information measures are totally cutoff-independent. This regulator involves cutting off geodesic curves at a horocycle, a geometric object defined in hyperbolic spaces with one point on the boundary, perpendicular to all geodesic curves approaching the boundary point. In this scheme, the entanglement entropy is precisely
\begin{equation}
    S(A) = 2 \ln \frac{L}{\epsilon} \,,
\end{equation}
without any corrections, and all cutoff dependence then cancels in finite information measures. The horocycle cutoff is thus a scheme in which no finite renormalization is required as one cutoff is replaced with another, and it is unique.

We generalize beyond the mutual information and demonstrate that for a broad class of information measures, finiteness under a general regulator implies total cutoff-independence with the horocycle regulator. For simple linear combinations of entropies, the finite measures are precisely the multipartite informations $I_k(A_1 : A_2 : \cdots : A_k) $ and their linear combinations, which include the conditional mutual information. In addition, a class of optimized correlation measures -- information measures that involve optimization over all possible purifications, and are monotonic under processing -- including quantities such as the entanglement of purification and the squashed entanglement, are argued to be in this class as well. 

It is natural to inquire what kind of regulator on the field theory side corresponds to the horocycle scheme. The horocycle regulator cuts off curves headed to a particular boundary point at different distances from the boundary depending on the angle of approach, which in turn is determined by the location of the far end of the spatial region. Thus on the field theory side, the horocycle regulator is non-local, providing a short-distance cutoff at a boundary point whose size scales with the distance to the far boundary point of the region. Important next questions concerning this work involve a better understanding of this non-local cutoff, and the generalization of the horocycle cutoff to higher dimensions. The horocycle regulator is naturally defined for one-dimensional curves in hyperbolic space, while in higher-dimensional cases, a regulator for surfaces is required.

The remainder of this paper is organized as follows. Section \ref{sec:horocycle} introduces the horocycle cutoff scheme on the Poincar\'e half-plane and compares it with the uniform radial cutoff, another standard regulator. Section \ref{sec:finite} will demonstrate how finite information measures are made cutoff independent by the horocycle regulator, and gives a set of conditions for information measures to be finite. Section \ref{sec:disk}  extends the analysis of the previous sections to another representation of two-dimensional hyperbolic space, the Poincar\'e disk. Section \ref{sec:CFT}  discusses properties of the CFT regulator dual to the bulk horocycle cutoff. Finally, we give concluding remarks in section \ref{sec:conc}. Other work on horocycles in the context of AdS/CFT has been done by L\'evay \cite{levay_berry_2019}.

\section{The horocycle cutoff} \label{sec:horocycle}

\begin{figure}
    \centering
    \includegraphics[width=0.6\linewidth]{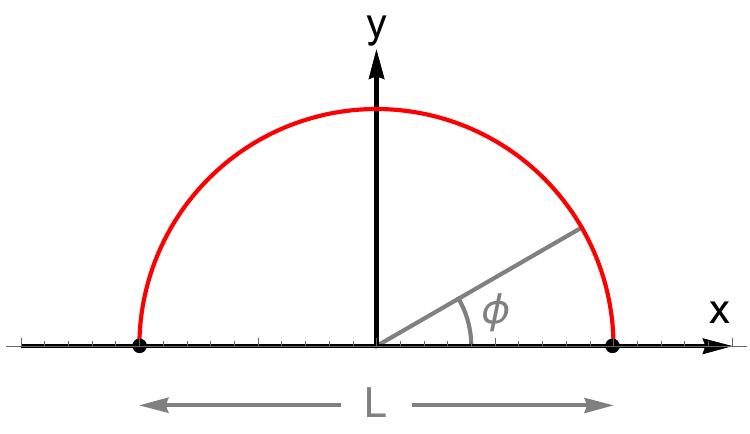}
    \caption{An RT surface (red) homologous to a boundary subregion of length $L$ in the Poincar\'e half-plane, representing a spatial slice  of an AdS$_3$. In two dimensions an RT surface is a geodesic, taking the form of the arc of a circle perpendicular to the boundary, here parametrized by the angular coordinate $\phi$.}
    \label{fig:HalfPlaneRT}
\end{figure}

We specialize to the vacuum of AdS$_3$/CFT$_2$, where spatial slices of the CFT are one-dimensional and spatial slices of the holographic bulk are two-dimensional hyperbolic space. One useful presentation of two-dimensional hyperbolic space is as the upper half-plane $\{x, y_{\geq 0}\}$ with the boundary at $y=0$ and metric
\begin{equation} \label{eq:HalfPlane_xy}
    ds^2 = \frac{dx^2 + dy^2}{y^2} \,.
\end{equation}
In this space, minimal area surfaces are geodesic curves, which take the form of arcs of circles intersecting the boundary at right angles. Thus for a one-dimensional interval of length $L$, the entanglement entropy of the CFT in that interval is just the area of the semi-circle extending into the bulk from the ends of the interval (see figure~\ref{fig:HalfPlaneRT}).

To calculate the length of the geodesic, it is convenient to use polar coordinates
\begin{equation}
    x = r \cos \phi \,, \quad \quad y = r \sin \phi \,,
\end{equation}
such that the metric takes the form
\begin{equation}
    ds^2 = \frac{dr^2 + r^2 d\phi^2}{r^2 \sin^2 \phi} \,.
\end{equation}
The naive entanglement entropy, calculated as the length of the geodesic, is then
\begin{eqnarray}
        S(A) = 2 \int_0^{\pi/2} \frac{d \phi}{\sin \phi} 
        = 2 \ln \tan \frac{\phi}{2} \Big|_0^{\pi/2} 
         = \infty \,,
\end{eqnarray}
which is divergent. This is a consequence of the boundary being infinitely far away.
Thus to define a notion of length for the geodesic, and hence for the entanglement entropy, we must regularize the calculation with some kind of cutoff. If we cut off the integral at an angle $\phi_0$, we find
\begin{eqnarray}
\label{eq:CutoffGeodesicLength}
    S(A) = 2 \int_{\phi_0}^{\pi/2} \frac{d \phi}{\sin \phi} 
        = 2 \ln \cot \frac{\phi_0}{2} \,.
\end{eqnarray}
To complete the calculation, we must define a prescription for choosing $\phi_0$ for any size region $L$.

Let us first consider a {\em uniform radial cutoff}. This means for calculating the geodesic curve for any region, we cut off the curve as it approaches the boundary at a fixed radial coordinate,
\begin{eqnarray}
    y = \epsilon  \,.
\end{eqnarray}
This corresponds to regulating $\phi$ at 
\begin{eqnarray}
    \phi_0 = \sin^{-1} \frac{2\epsilon}{L}\,,
\end{eqnarray}
and for the geodesic length we obtain
\begin{eqnarray}
       S(A)  &=& 2 \ln \cot \frac{\sin^{-1}  (2\epsilon /L)}{2} \\
       &=& 2\, {\rm sech}^{-1} \frac{2\epsilon}{L} \,,
       \label{eq:Entropy_Uniform_Radial_Cutoff}
\end{eqnarray}
 where the second line follows from the identity sech $\!\ln \cot x = \sin 2x$.
As the cutoff surface becomes close to the boundary, we can expand in small $\epsilon/L$ and we find
\begin{equation}
    S(A)=  2 \ln \frac{L}{\epsilon} - \frac{2\epsilon^2}{L^2} + {\cal O}(\epsilon^4) \,.
\end{equation}
This has the form of a leading logarithmically divergent part, as well as subleading terms that vanish as the cutoff is removed by $\epsilon \to 0$. Thus if we keep the terms that are diverging or finite in the $\epsilon \to 0$ limit, and discard those that vanish, we have as an entanglement entropy for region $A$ with length $L$,
\begin{eqnarray}
\label{eq:EntropyRadialCutoff}
    S(A) \to  2 \ln \frac{L}{\epsilon} \,.
\end{eqnarray}
However, for any finite value of the cutoff this is modified by the ${\cal O}(\epsilon^2)$ correction terms. Other generic cutoffs will have similar behavior; for example, cutting off at angle $\phi_0 = 2\epsilon/L$ also leads to (\ref{eq:EntropyRadialCutoff}) plus ${\cal O}(\epsilon^2)$ terms that vanish as we remove the cutoff, though these vanishing terms are different.

\begin{figure}
    \centering
    \includegraphics[width=0.6\textwidth]{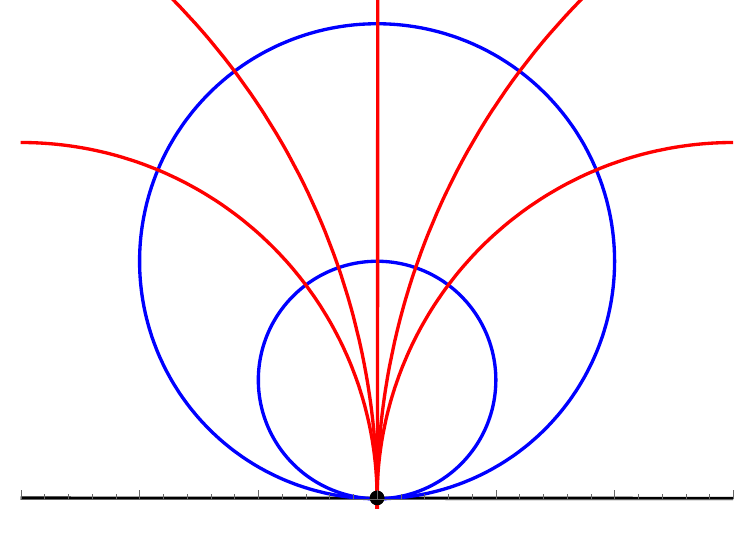}
    \caption{Two concentric horocycles (blue curves) with ``center" at the origin (black dot). All geodesics converging to the center (red curves) pass perpendicularly through the concentric horocycles.}
    \label{fig:ConcentricHorocycles}
\end{figure}

Now consider the {\em horocycle cutoff}. A horocycle in hyperbolic space is a curve whose perpendicular geodesics all converge to a single boundary point, called the ``center". In our coordinate system, a horocycle takes the form of a circle tangent to a single point on the boundary, the boundary point being the center (note that this is not the coordinate center of the circle) to which all perpendicular geodesics converge. Horocycles of different sizes sharing a center are called ``concentric", and all geodesics approaching the center of one horocycle will also pass perpendicularly through all concentric horocycles (see figure~\ref{fig:ConcentricHorocycles}).

\begin{figure}
    \centering
    \includegraphics[width=0.45\textwidth]{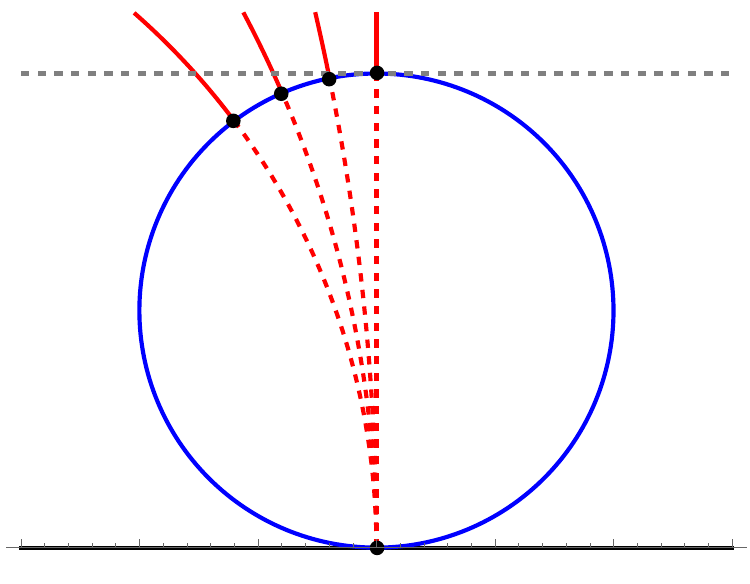}
    \hskip0.5in 
    \includegraphics[width=0.45\textwidth]{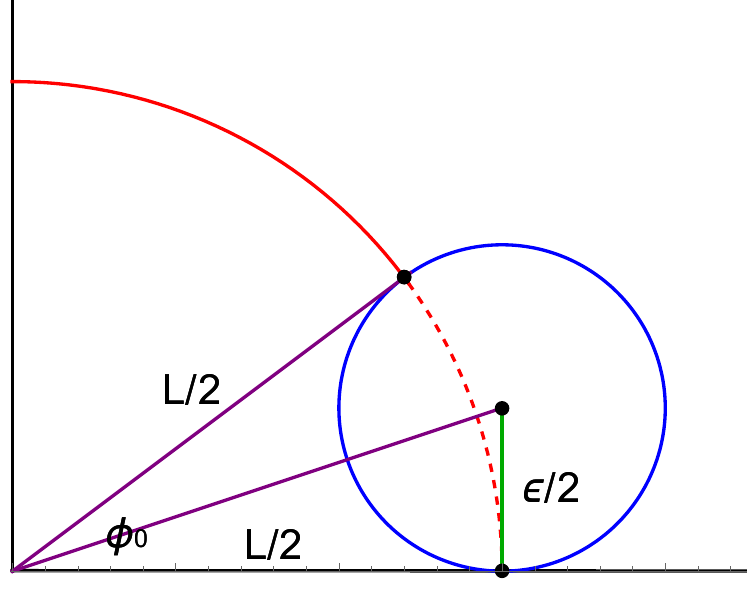}
    \caption{The horocycle cutoff. Left: different geodesics converging to the same boundary point are each cut off where they intersect a horocycle of fixed diameter $\epsilon$. This differs from a uniform radial cutoff, which would cut them all off at the dashed gray line. Right: a geodesic circle of diameter $L$ is cut off at angle $\phi_0$ from the horizontal where $\tan \phi_0/2 = \epsilon/L$.}
    \label{fig:HorocycleCutoff}
\end{figure}

The idea of a horocycle cutoff is that instead of picking a fixed radial cutoff, one picks a horocycle of fixed size; let us call the horocycle diameter $\epsilon$. Then any geodesic approaching a given point on the boundary will be cut off when it intersects the horocycle with diameter $\epsilon$ whose center is the boundary point being approached (see the left of figure~\ref{fig:HorocycleCutoff}). Different geodesics are then cut off at different radial distances from the boundary, with the perpendicular geodesic cut off at radial distance $\epsilon$ but geodesics approaching at other angles cut off closer to the boundary.

Let us reconsider our calculation of the length of a cutoff geodesic, and hence of the entanglement entropy. A geodesic that is an arc of a circle with diameter $L$ will be cut off at an angle $\phi_0$ obeying
\begin{eqnarray}
\label{HorocycleAngleCutoff}
    \tan \frac{\phi_0}{2} = \frac{\epsilon}{L} \,,
\end{eqnarray}
as can be seen geometrically from the right-hand-side of figure~\ref{fig:HorocycleCutoff}. 
We can view the horocycle cutoff as an effective radial cutoff that depends on $L$, the size of the boundary region enclosed by the geodesic. Defining the effective radial cutoff as
\begin{eqnarray}
    \epsilon_{\rm eff} \equiv \frac{L}{2} \sin \phi_0 \,,
\end{eqnarray}
we find using the identity $\sin 2x = 2 \tan x/(1 + \tan^2 x)$ that for the horocycle cutoff, each geodesic is terminated at radial distance
\begin{eqnarray}
\label{HorocycleEffectiveCutoff}
 y(L)=     \epsilon_{\rm eff} = \frac{\epsilon}{1 + (\epsilon/L)^2}  \,,
\end{eqnarray}
which in the limit of large regions approaches $\epsilon$, but becomes smaller as the size of the region decreases.

We see immediately upon using the horocycle prescription for the regulated form of the entanglement entropy (\ref{eq:CutoffGeodesicLength}) that the result is exactly
\begin{eqnarray}
\label{eq:ExactEntropy}
    S(A) = 2 \ln \frac{L}{\epsilon} \,.
\end{eqnarray}
This coincides with the divergent and finite pieces of the result from the uniform radial cutoff (\ref{eq:EntropyRadialCutoff}), but for a finite value of the cutoff this expression receives no corrections at any order in $\epsilon$, and is instead exact.  We can see that the horocycle cutoff formula (\ref{HorocycleAngleCutoff}) -- or equivalently (\ref{HorocycleEffectiveCutoff}) -- is the unique cutoff that results in the geodesic length (\ref{eq:CutoffGeodesicLength}) evaluating to exactly (\ref{eq:ExactEntropy}); even if we didn't know the geometric interpretation of the horocycle, we could choose (\ref{HorocycleAngleCutoff}) purely on the basis of being the necessary expression to give us (\ref{eq:ExactEntropy}) without any corrections subleading in $\epsilon$.

The horocycle prescription is well-defined for calculating the lengths of curves, as needed for the holographic entanglement entropy. For a different calculation, such as an integral over all bulk space of the sort needed to calculate a correlation function of local operators, the horocycle cutoff lacks a natural definition. An ordinary radial cutoff could be employed in such a case; whether there are natural generalizations for the horocycle cutoff to integrals over surfaces of different dimension is a question we return to in the conclusions.

We shall learn more about the special properties of horocycles in the next section, where we consider quantities that are finite in the limit the  cutoff is removed, and see how the horocycle prescription renders them entirely independent of the cutoff.

\section{Finite information measures and total cutoff independence} \label{sec:finite}

While the entanglement entropy of a subregion is divergent in the absence of a cutoff, there are other information measures built from entropies that are finite in the limit. We will start by discussing the most familiar of these, the mutual information $I(A:B)$ of two regions $A$ and $B$. We shall show that, while any regulator produces the same value for the mutual information when the cutoff is removed, the horocycle regulator achieves the correct value for finite values of the cutoff as well. Later in the section, we will describe broader classes of information measures with this same property.

\subsection{Mutual information}
\begin{figure}
    \centering
    \includegraphics[width=0.6\linewidth]{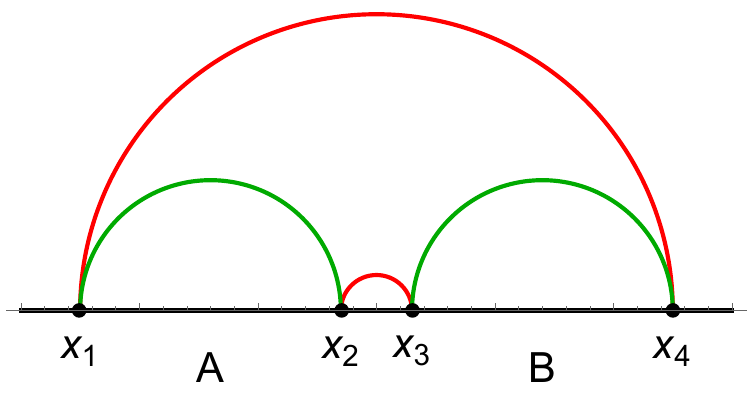}
    \caption{The mutual information $I(A:B)$ for two boundary subregions $A$ and $B$ in the Poincar\'e half-plane. Green RT surfaces contribute positively to the mutual information, while red RT surfaces contribute negatively.}
    \label{fig:MI}
\end{figure}

Given two disjoint subregions $A$ and $B$, the mutual information is defined in terms of the joint entropy $S(AB)$ and the marginal entropies $S(A)$ and $S(B)$ as
\begin{eqnarray}
    I(A:B) = S(A) + S(B) - S(AB) \,.
\end{eqnarray}
Let us consider region $A$ stretching from $x_1$ to $x_2$ and region $B$ stretching from $x_3$ to $x_4$, with $x_4 > x_3 > x_2 >x_1$. $S(A)$ and $S(B)$ always correspond to the lengths of geodesics stretching from $x_1$ to $x_2$, and $x_3$ to $x_4$, respectively. Depending on the relative sizes of the various regions, $S(AB)$ will either be the sum of the $A$ and $B$ geodesics or, if it is shorter, the sum of the lengths of the geodesics stretching from $x_2$ to $x_3$ and from $x_1$ to $x_4$ (see figure~\ref{fig:MI}). In the former case, the mutual information is zero. Let us assume the latter case and calculate this quantity.

In the uniform radial cutoff prescription, we have from (\ref{eq:Entropy_Uniform_Radial_Cutoff}),
\begin{eqnarray}
\label{eq:RadialMutualInfo}
    I(A:B) = 2\left( {\rm sech}^{-1} \frac{2\epsilon}{x_2 - x_1} +    {\rm sech}^{-1} \frac{2\epsilon}{x_4 - x_3} - 
       {\rm sech}^{-1} \frac{2\epsilon}{x_3 - x_2} -    {\rm sech}^{-1} \frac{2\epsilon}{x_4 - x_1} \right) \,,
\end{eqnarray}
which for small cutoff becomes
\begin{eqnarray}
\label{eq:RadialMutualInfoLimit}
    I(A:B) = 2 \ln \frac{(x_2 - x_1) (x_4-x_3)}{(x_3-x_2)(x_4-x_1)} - 2\epsilon^2  \left( \frac{1}{(x_2-x_1)^2} + \frac{1}{(x_4-x_3)^2} - \frac{1}{(x_3-x_2)^2} - \frac{1}{(x_4-x_1)^2}  \right)  + {\cal O}(\epsilon^4) \,.
\end{eqnarray}
In the $\epsilon \to 0$ limit, this converges to a finite value,\footnote{It is precisely when this quantity is non-negative that the prescription for calculating $S(AB)$ was correct; if this quantity is negative, the correct prescription was $S(AB) = S(A) + S(B)$ and $I(A:B) = 0$.}
\begin{eqnarray}
    I(A:B) \to 2 \ln \frac{(x_2 - x_1) (x_4-x_3)}{(x_3-x_2)(x_4-x_1)}  \,.
\end{eqnarray}
Thus the mutual information is well-defined in the limit the cutoff is removed; it is {\em finite}. For any nonzero value of the cutoff, however, the result (\ref{eq:RadialMutualInfo}) receives corrections as given in  (\ref{eq:RadialMutualInfoLimit}).

Consider now the horocycle cutoff prescription. Using (\ref{eq:ExactEntropy}) we find
\begin{eqnarray}
    I(A:B) &=& 2\left( \ln \frac{x_2 - x_1}{\epsilon } +    \ln \frac{x_4 - x_3}{\epsilon } - 
       \ln \frac{x_3 - x_2}{\epsilon } -   \ln \frac{x_4 - x_1}{\epsilon }\right) \\
       &=& 2 \ln \frac{(x_2 - x_1) (x_4-x_3)}{(x_3-x_2)(x_4-x_1)} \,.
\end{eqnarray}
For any value of $\epsilon$ in the horocycle cutoff, the mutual information is $\epsilon$-independent and is exactly equal to its $\epsilon \to 0$ limit in the radial cutoff. So the horocycle cutoff gives the exact value for the mutual information at finite values of the cutoff, without taking any limit.

\begin{figure}
    \centering
    \includegraphics[width=0.5\textwidth]{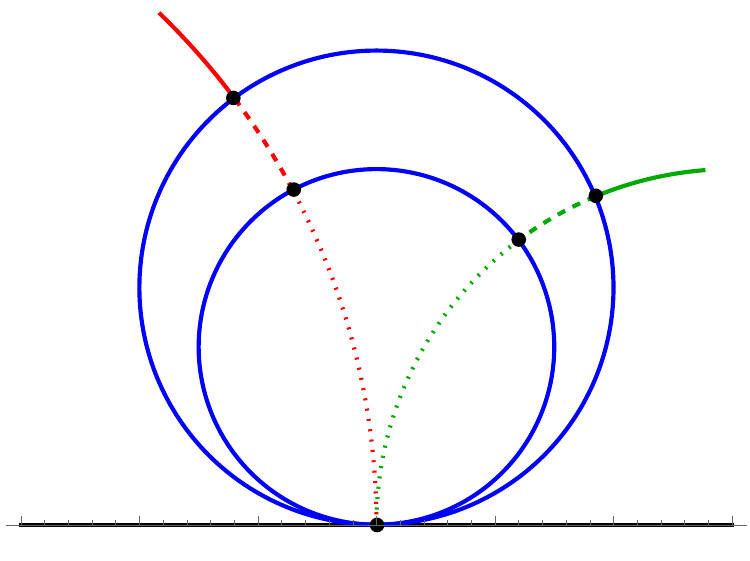}
    \caption{As the horocycle cutoff is changed, all geodesics change by the same length, indicated by the dashed portions of the red and green geodesics between the two concentric horocycles. Thus there is no net change to the mutual information or other finite information measure.}
    \label{fig:ConcentricCutoff}
\end{figure}

We can see geometrically how the horocycle prescription achieves this cutoff-independence. In the Ryu-Takayanagi prescription for entropy, the divergent parts come from a geodesic stretching to the boundary. A quantity like the mutual information achieves a finite result by balancing each geodesic approaching the boundary at a certain point with another geodesic approaching the same point, but entering the formula with opposite sign. This is indicated in the figure as a green line for a geodesic contributing with a positive sign, and a red line for a geodesic contributing with a negative sign, and we can see in figure~\ref{fig:MI} how for each boundary point $x_1$, $x_2$, $x_3$ and $x_4$ there is a positive and a negative geodesic approaching.

Consider a boundary point with two balancing geodesics approaching it, and imagine varying the cutoff. With a generic cutoff, such as the uniform radial prescription, each geodesic will incur a change in length that depends on the angle it approaches the boundary, and hence on the length of the boundary region associated to that geodesic. Since the balancing geodesics will in general be associated to boundary regions of different sizes, the change in length for each of the boundary geodesics will be different. Due to this imperfect cancellation there will be a  finite change in the net length as the cutoff is varied.

The horocycle, however, is exactly the prescription to avoid this. Consider two concentric horocycles with diameters $\epsilon_1$ and $\epsilon_2$ (see figure~\ref{fig:ConcentricCutoff}). For a single geodesic approaching the boundary point that is the ``center" of the horocycles, the change in its length between being cut off at one horocycle and being cut off at the other is
\begin{eqnarray}
    \Delta \ell  = 2 \ln \tan \frac{\phi_2}{2} - 2 \ln \tan \frac{\phi_1}{2} = 2 \ln \frac{\epsilon_2/L}{\epsilon_1/ L} = 2 \ln \frac{\epsilon_2}{\epsilon_1} \,,
\end{eqnarray}
which is independent of $L$, the size of the boundary region bounded by the geodesic. The change in length upon adjusting the horocycle cutoff is thus the same for {\em all} geodesics, and thus the balancing geodesics stay balanced and the finite value of $I(A:B)$ does not change.

Here we have compared the horocycle cutoff to the uniform radial cutoff, but a generic cutoff other than the horocycle cutoff will have subleading terms depending on a ratio of $\epsilon$  to the size of the region, which will not cancel for finite cutoff; the horocycle cutoff is the unique prescription that obtains the correct result for finite cutoff.

Before generalizing these considerations to other information measures, we point out an important special case. In the above consideration of the mutual information, we have assumed that the regions $A$ and $B$ do not share any boundary points. When they do, the mutual information becomes divergent. For example, if we take $x_3 = x_2$ and then calculate the mutual information with the horocycle regulator, we find
\begin{eqnarray}
    I(A:B) = 2\left( \ln \frac{x_2 - x_1}{\epsilon } +    \ln \frac{x_4 - x_2}{\epsilon } - \ln \frac{x_4 - x_1}{\epsilon }\right)  = 2 \ln \frac{(x_2 - x_1) (x_4-x_2)}{\epsilon(x_4-x_1)} \,,
\end{eqnarray}
which goes to infinity as we remove the cutoff; other regulators give the same result with additional subleading terms.  The limit $x_3 \to x_2$ is not smooth because taking $A$ and $B$ to share a boundary eliminates the geodesic that stretched between the two boundary points, but this term was divergent in the limit, so removing it is not the same as including it and taking the $(x_3 - x_2) \to 0$ limit; instead the proper calculation cuts the limit off at $(x_3 - x_2) \to \epsilon$. Physically, this corresponds to the divergent entanglement across the shared boundary of the two regions. 

\subsection{Linear entropic formulas}
The mutual information for separated regions is {\em finite} for any reasonable regulator, and we have shown that with the horocycle regulator, it obeys {\em complete cutoff independence}. It is natural to identify other finite information measures that are fully independent of the cutoff using the horocycle regulator.
As discussed in the last subsection, the finiteness owes to each diverging geodesic approaching a given boundary point being balanced by another geodesic approaching the same point  weighted with opposite coefficient in the formula for $I(A:B)$. This generalizes to a number of other information measures.

For a quantum system containing $n$ subsystems or ``parties" $A_i$ (which for us will be boundary regions or sets of boundary regions), there are $2^n - 1$ distinct entropies $S(A_1 A_2 \cdots A_k)$ that can be built from $k \leq n$ parties; it is common to think of these entropies as comprising a basis for a vector space, and inequalities like subadditivity $S(A_1) + S(A_2) \geq S(A_1A_2)$ and its generalizations constrain the physically allowed vectors to a  
cone, the {\em entropy cone}.\footnote{The {\em holographic} entropy cone, applying to systems with a classical gravity dual, is more constrained by relations such as the monogamy of mutual information \cite{Hayden:2011ag} as well as additional relations for more parties \cite{Bao:2015bfa}; for further references see for example \cite{Hernandez-Cuenca:2023iqh}.}  A vector in this space corresponds to a linear combination of entropies, or {\em linear entropic formula}; the mutual information $I(A_1:A_2)$ is an example. 

A linear entropic formula on $k$ parties (spatial regions) will be finite if the set of geodesics terminating on each boundary point is balanced in its contribution to the Ryu-Takayanagi formula. When the boundaries of the spatial regions $A_i$ are disjoint, we have
\begin{itemize}
    \item A linear entropic formula is finite when, for each party $A_i$, the coefficients of each entropy containing $A_i$ sum to zero.
\end{itemize}
The property was called {\em balanced} in \cite{Hubeny:2018trv}. It is straightforward to find a basis of quantities satisfying this condition. The {\em $k$-party multipartite informations}  $I_k$ can be defined
\begin{eqnarray}
\label{eq:MultipartiteInformation}
   I_k(A_1 : A_2 : \cdots : A_k) \equiv \sum_{j =1}^k \left( (-1)^{j-1} \sum_{i_1 < i_2 < \ldots i_j} S(A_{i_1} A_{i_2} \ldots A_{i_j}) \right)\,,
\end{eqnarray}
which simply sums all the entropies up to $k$ parties with plus signs for an odd number of parties and minus signs for  an even number of parties.  The multipartite informations are totally symmetric under the exchange of all $k$ parties, and it is easy to see that the $k=2$ case is the ordinary mutual information $I(A_1:A_2)$, and the $k=3$ case is the tripartite information,
\begin{equation} \label{eq:I3}
    I_3(A_1:A_2:A_3) \equiv S(A_1) + S(A_2) + S(A_3) - S(A_1A_2)-S(A_1A_3)-S(A_2A_3)+S(A_1A_2A_3).
\end{equation}
We can show that the $k$-party multipartite information for $k \geq 2$ is finite in the sense we have discussed; consider any one party, say $A_1$. The sum in (\ref{eq:MultipartiteInformation}) involving entropies with $j$ parties will have $\begin{pmatrix} k-1 \cr j-1 \end{pmatrix}$ terms including $A_1$. For finiteness the total coefficient of these terms must vanish, and indeed we have
\begin{eqnarray}
\label{eq:BinomialCoeff}
    \sum_{j = 1}^k (-1)^{j-1} \begin{pmatrix} k-1 \cr j-1 \end{pmatrix} =  \sum_{\ell=0}^{k-1} (-1)^\ell \begin{pmatrix} k-1 \cr \ell \end{pmatrix} = 0\,,
\end{eqnarray}
an identity for binomial coefficients with $k \geq 2$. (It does not hold for $k=1$, which is just the statement that the ordinary entropies $S(A_i)$ are not finite.) The $I_k$ were shown to be balanced in \cite{Hubeny:2018trv}.

Thus the multipartite informations $I_k$ with $k \geq 2$ are all finite. Moreover we can see that any finite linear entropic formula can be written as a linear combination of the $I_k$. To see this, recall that the $2^n - 1$ entropies $S(A_i)$, $S(A_i A_j)$, \ldots $S(A_1 A_2 \ldots A_n)$ form a basis for the $n$-party entropy cone. However, we can clearly substitute $I_n(A_1 : \cdots : A_n)$ for $S(A_1 A_2 \ldots A_n)$ as a basis vector, and then we can substitute the $n$ distinct $I_{n-1}$ for the $n$ entropies with all but one of the parties, and so on, until we obtain as the basis for the entropy cone all the single-party entropies $S(A_i)$ and the multipartite informations $I_k$, $2 \leq k \leq n$.

There are $2^n - 1 -n$ such $I_k$. The entropy cone is $2^n - 1$ dimensional, and the conditions for finiteness impose one condition for each of the $n$ parties on the linear entropic formula; thus the space of finite linear entropic formulas is also has dimension $2^n - 1 -n$. Thus the multipartite informations provide a basis for finite linear entropic formulas (for more discussion on this basis, see \cite{He:2019ttu}). They are fully cutoff-independent when used with the horocycle regulation scheme.

As an example, for $n=3$ with parties $A$, $B$, and $C$ we have a seven-dimensional entropy cone with basis $S(A)$, $S(B)$, $S(C)$, $S(AB)$, $S(AC)$, $S(BC)$, and $S(ABC)$. An alternate basis is the three entropies $S(A)$, $S(B)$, and $S(C)$, the mutual informations $I(A:B)$, $I(A:C)$, and $I(B:C)$, and the tripartite information $I_3(A:B:C)$ given in (\ref{eq:I3}).  Thus we have that the mutual informations and the tripartite information are all finite, and moreover are cutoff independent with the horocycle regulator. Moreover, linear combinations of these are finite as well, such as the mutual information between one party and the other two,
\begin{eqnarray}
    I(A:BC) = I(A:B) + I(A:C) - I_3(A:B:C) \,,
\end{eqnarray}
and the conditional mutual information,
\begin{eqnarray}
    I(A:B|C) = I(A:B)  - I_3(A:B:C) \,.
\end{eqnarray}
Thus we see, for any number of parties, any linear combination of the $I_k$ will be finite in the sense of having a finite limit as the regulator is removed, and moreover, will be fully regulator-independent for nonzero regulator when the horocycle cutoff is used. We note that the $I_k$ in general do not have definite sign, so an individual one of these quantities may not serve as a measure of total correlation, but their finite/balanced property holds regardless.

The above discussion applies when the regions $A_i$ do not share any mutual boundary points. We found in the last subsection that if two regions share a boundary point, the mutual information depends on the cutoff, and is divergent. The appropriate generalization of the finiteness condition is
\begin{itemize}
    \item A linear entropic formula remains finite when two regions $A_i$ and $A_j$ share a boundary point if the coefficients of all entropies containing $A_i$ or $A_j$, but not both, sum to zero.
\end{itemize}
It is straightforward to show, using a generalization of the arguments leading to (\ref{eq:BinomialCoeff}), that the $I_k$, $k \geq 3$ remain finite when one or more boundary points is shared between regions. Thus of all the $I_k$ with $k \geq 2$, only the mutual information becomes divergent when boundaries coincide (at least for the case of two-dimensional field theories). The $I_k$, $k \geq 3$ have been termed {\em superbalanced}, and have the property that if one party is exchanged with the purifier they remain balanced \cite{Hubeny:2018ijt,He:2020xuo}; the fact that the superbalanced measures remain finite when a boundary region coincides was noted in \cite{Hubeny:2018ijt, Avis:2021xnz}, see \cite{Hernandez-Cuenca:2023iqh} for a review.

In general if a region $A_i$ has both shared and unshared boundary points, both the original condition on $A_i$ alone as well as the condition on $A_i$ with its neighbor region $A_j$ must hold separately to avoid divergences.

\subsection{Optimized correlation measures}
Not all information measures are simple linear entropic formulas. Another class is the {\em optimized} information measures, where the quantity is defined as the extremum of a formula over all possible extensions or purifications of the quantum state being considered. For example, the entanglement of purification \cite{EP02} is defined as
\begin{eqnarray}
      E_P(A:B) \equiv \inf_{|\psi\rangle} S(Aa) \,,
\end{eqnarray}
where the infimum is over all possible states $|\psi\rangle_{ABab}$ purifying $\rho_{AB}$; $E_P$ has the operational meaning of the entanglement cost of creating a state asymptotically from Bell pairs with negligible communication.  Another example is the squashed entanglement \cite{tucci_entanglement_2002, christandl_squashed_2004},
\begin{eqnarray}
    E_{\rm sq}(A:B) \equiv \inf_{|\psi\rangle} I(A:B|a)\,,
\end{eqnarray}
which vanishes on separable states and thus measures only quantum entanglement, not classical correlations. 

\begin{figure}
    \centering
    \includegraphics[width=0.3\linewidth]{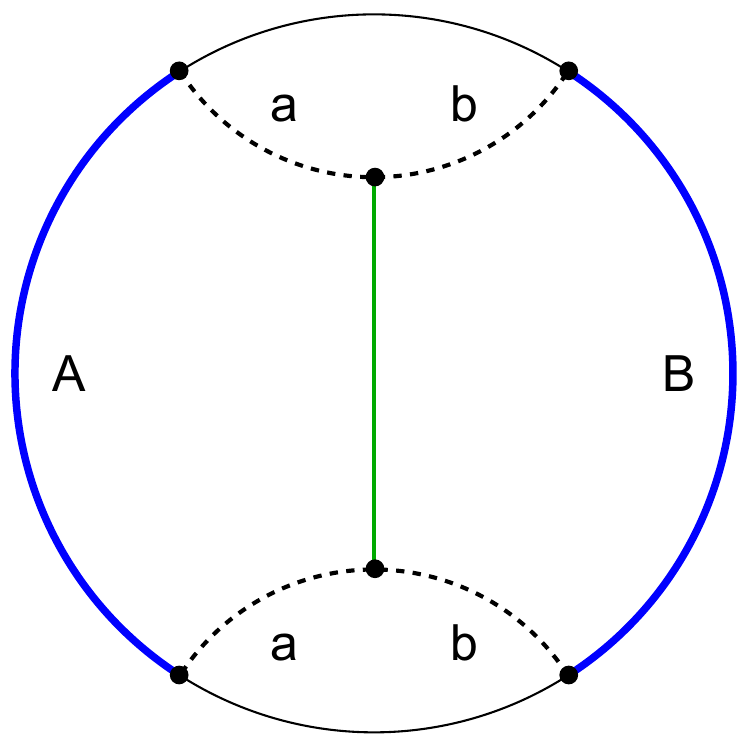}
    \hspace{2cm}
    \includegraphics[width=0.3\textwidth]{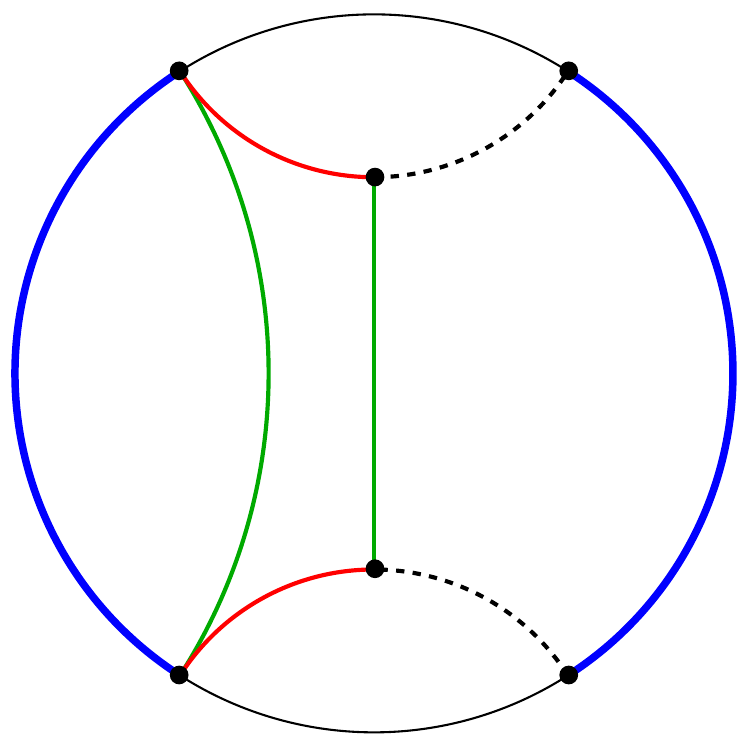}
    \caption{The holographic duals of the optimized correlation measures $E_P$ (left) and $E_Q$ (right) in the Poincar\'e disk. Green curves contribute positively to the quantity, while red curves contribute negatively. Figures reproduced from \cite{dewolfe_multipartite_2020}.}
    \label{fig:EP_EQ}
\end{figure}

An interesting subclass is the {\em optimized correlation measures}, optimized information measures that also obey monotonicity, meaning when part of a party is traced out or otherwise processed, the measure can never increase:
\begin{eqnarray}
    E(A:B) \leq E(AA':B) \,.
\end{eqnarray}
$E_P$ and $E_{\rm sq}$ are both optimized correlation measures, and  optimized correlation measures on two parties were classified in \cite{Levin_2020}, where the additional measures $Q$-correlation $E_Q$ and $R$-correlation $E_R$ were discovered. Gravity duals for $E_P$ and $E_{\rm sq}$ were proposed first, the former being calculated by the entanglement wedge cross-section\footnote{There have also been proposals that the entanglement wedge cross-section holographically calculates the {\em reflected entropy} \cite{dutta_canonical_2021}; it is possible that these quantities coincide in the limit of classical gravity. Any information measure calculated by the entanglement wedge cross-section is finite since the surface does not anywhere approach the boundary.} \cite{Takayanagi:2017knl,Nguyen:2017yqw} and the later coinciding with the mutual information in holographic states with classical gravity duals \cite{Umemoto:2018jpc}, while duals for $E_Q$ and $E_R$ were proposed in \cite{Levin:2019krg}, the latter coinciding with $E_P$. Generalizations to black hole geometries were studied in \cite{Agrawal:2021nkw}, where again it was found that the holographic dual of $E_R$ coincided with $E_P$, while a rich structure of phase transitions was uncovered for the holographic dual of $E_Q$. Figure \ref{fig:EP_EQ} depicts the holographic duals of $E_P$ (equivalently, $E_R$) and $E_Q$ in the Poincar\'e disk, where again green curves contribute positively and red curves contribute negatively. All these quantities are finite in the sense we have discussed here; for measures calculated by the entanglement wedge cross section there are no geodesics going to the boundary, so the result should be cutoff independent in any regulator, while for $E_{\rm sq}$ (which coincides with the mutual information) and $E_Q$, curves always approach boundary points in balanced pairs. 

In \cite{dewolfe_multipartite_2020}, a classification of tripartite symmetric optimized correlation measures was undertaken. There it was also argued that any such correlation measure will be finite\footnote{The terminology used there was {\em cutoff-independent}, but the meaning was what is called finite here. The measures are fully cutoff-independent with the horocycle regulator.} in the sense we describe, where geodesics approaching the boundary contributing to the information measure are balanced by other geodesics. 
Using the language described here, this finiteness can be understood as follows. The prescription for the gravity dual involves purifications that extend into the bulk. Each party $A_i$ is associated with a set of degrees of freedom in the extension $a_i$, and it is assumed that geometrically each $a_i$ has part of its boundary on the boundary of $A_i$, while the rest of the boundary of $a_i$ is in the bulk; all of the boundary of $A_i$ is covered in this way. Thus the boundary points shared between $A_i$ and $a_i$ require that the sum of coefficients of all terms involving $A_i$ or $a_i$ sum to zero; no separate constraint is placed on $a_i$ alone since its other boundary points are in the bulk. But it was shown in  \cite{dewolfe_multipartite_2020} that the monotonicity property of optimized correlation measures follows if  all terms in the optimization function for the correlation measure involving $A_i$ or $a_i$ satisfy one of:
\begin{enumerate}
    \item The entropy in the term is independent of $A_i$ and $a_i$ or contains both $A_i a_i$.
    \item A pair of terms is of the form $S(A_iX) - S(a_i Y)$ where $X$ and $Y$ are disjoint and contain neither $A_i$ nor $a_i$.
\end{enumerate}
Terms satisfying either of these rules will always satisfy the finiteness conditions outlined above, the former because no associated curves approach the boundary, and the latter because they always approach the boundary in balanced pairs. Thus, optimized correlation measures are also finite as the cutoff is removed, and can be made independent of a finite cutoff if the horocycle regulator is used.

\section{The horocycle cutoff for the Poincar\'e disk} \label{sec:disk}

\begin{figure}
    \centering
    \includegraphics[width=0.4\linewidth]{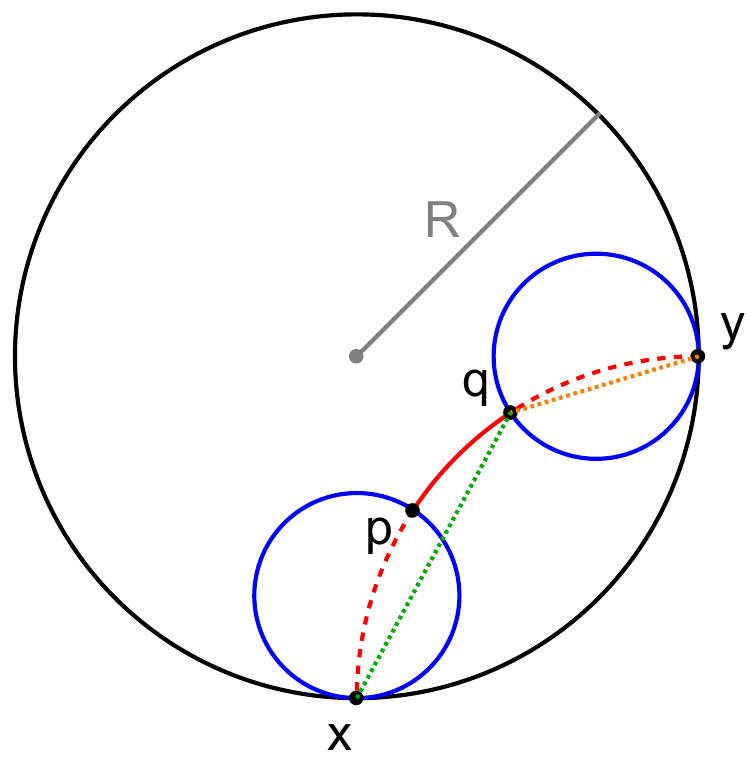}
    \caption{A Poincar\'e disk of radius $R$ with two horocycles (blue) of the same size regulating an RT surface (red). The size of the horocycles is exaggerated for visual clarity.}
    \label{fig:disk_horocycles}
\end{figure}

The total cutoff independence provided by horocycles is not unique to the Poincar\'e half-plane, but holds also for the 2D Poincar\'e disk. Consider the metric
\begin{equation} \label{eq:disk_metric}
    ds^2 = \frac{4}{(R^2-r^2)^2} \left( dr^2 + r^2 d\theta^2 \right)\,,
\end{equation}
as a time slice of AdS$_3$. Here we have introduced the parameter $R$ as the radius of the Poincar\'e disk; we could set this to unity but will find it useful to keep it explicit. This metric is related to the Poincar\'e half-plane by the Cayley transformation, and to the global AdS metric with boundary at infinity by the coordinate transformation
\begin{equation}
    \frac{r}{R} = \tanh \frac{\rho}{2}\,.
\end{equation}
In this geometry, geodesics still take the form of arcs of circles intersecting the boundary at right angles, and horocycles are still Euclidean circles tangent to a single point on the boundary at $r=R$. As before, we regulate the area of an RT surface with horocycles placed at each endpoint on the boundary; see figure~\ref{fig:disk_horocycles} for an example. The intersection between the RT surface and horocycles is used to cut off the RT surface. This can be used to define an effective radial cutoff similar to (\ref{HorocycleEffectiveCutoff}) for the Poincar\'e half-plane; in the Poincar\'e disk, the effective radial cutoff is given by
\begin{equation} \label{eq:effUV_disk}
    \epsilon_\text{eff, disk} = R \left[ 1 - \left( 1 + \frac{4\epsilon(\epsilon - 2R)}{4R^2 + \epsilon^2\cot^2(L/2R)} \right)^{1/2} \right]\,.
\end{equation}
We note that this expression matches (\ref{HorocycleEffectiveCutoff}) for the Poincar\'e half-plane in the $R \rightarrow \infty$ limit.
The length of this regulated RT surface is most readily computed using the cross-ratio. Consider an ordered list of points $\{x,p,q,y\}$ along a geodesic with points $x$ and $y$ located on the boundary; see figure~\ref{fig:disk_horocycles}. The geodesic length between points $p$ and $q$ is given by
\begin{equation} \label{eq:geo_length}
    d(p,q) = \log \frac{|xq| |py|}{|xp| |qy|}\,,
\end{equation}
where $|\ldots|$ denotes Euclidean distance. Choosing points $p$ and $q$ to be the intersections between the RT surface and the two horocycles (both with diameter $\epsilon$), we find
\begin{align}
    |xq| &= |py| = 2R' \sin\bigg[ \arctan\left(\frac{R}{R'}\right) - \arctan\left(\frac{\epsilon}{2R'}\right) \bigg]\,, \\
    |xp| &= |qy| = 2R' \sin\bigg[ \arctan\left(\frac{\epsilon}{2R'}\right) \bigg]\,,
\end{align}
where
\begin{equation}
    R' = R \tan \left( \frac{L}{2R} \right)\,,
\end{equation}
and $L$ is the length of the boundary subregion. The lengths $|xq|$ and $|qy|$ are marked as green and orange dashed lines (respectively) in figure~\ref{fig:disk_horocycles}. Plugging these lengths into (\ref{eq:geo_length}), we find the horocycle regulated entropy of the subregion,
\begin{equation} \label{eq:ExactEntropy_Disk}
    S(A) = 2 \ln \left[ \sin\left(\frac{L}{2R}\right) \left(\frac{2R}{\epsilon} - 1\right) \right]\,,
\end{equation}
which as expected is divergent as $\epsilon \to 0$.
Again, this expression matches (\ref{eq:ExactEntropy}) for the Poincar\'e half-plane in the $R \rightarrow \infty$ limit.

Let us now consider the mutual information of two disconnected subregions $A$ and $B$ on the boundary of the Poincar\'e disk. Let $C$ and $D$ be the disconnected subregions forming the complement of $AB$, and let $L_i$ denote the length of subregion $i$. Using (\ref{eq:ExactEntropy_Disk}), the mutual information of $A$ and $B$ is given by
\begin{align*}
    I(A:B) &= 2 \ln \left[ \sin\left(\frac{L_A}{2R}\right) \left(\frac{2R}{\epsilon} - 1\right) \right] + 2 \ln \left[ \sin\left(\frac{L_B}{2R}\right) \left(\frac{2R}{\epsilon} - 1\right) \right] \\
        &- 2 \ln \left[ \sin\left(\frac{L_C}{2R}\right) \left(\frac{2R}{\epsilon} - 1\right) \right] - 2 \ln \left[ \sin\left(\frac{L_D}{2R}\right) \left(\frac{2R}{\epsilon} - 1\right) \right] \numberthis \\[0.3cm]
    &= 2 \ln \frac{\sin(\frac{L_A}{2R})\sin(\frac{L_B}{2R})}{\sin(\frac{L_C}{2R})\sin(\frac{L_D}{2R})}. \numberthis
\end{align*}
All $\epsilon$-dependence has completely canceled, leaving the mutual information totally cutoff independent. One can see that this will generalize to other finite information measures, since the $\epsilon$-dependent term in (\ref{eq:ExactEntropy_Disk}) can be cast as an $L$-independent additive constant, which will cancel out when there are balancing geodesics approaching the same boundary point. Therefore, horocycles also provide a totally cutoff-independent regularization of finite information measures for the Poincar\'e disk.

\section{Field theory interpretation of the horocycle cutoff} \label{sec:CFT}

The horocycle cutoff is a bulk regulator, but any bulk regulator implies a regulator scheme for the dual field theory via the AdS/CFT correspondence. Here, we comment on the interpretation of the horocycle regulator in the dual boundary CFT. 

Let us  first remind ourselves of the CFT interpretation of the uniform radial cutoff. Anti-de Sitter space has an isometry that is a combination of a shift in the radial direction and a scale transformation in the dual field theory. In the two-dimensional Poincar\'e half-plane with metric given by (\ref{eq:HalfPlane_xy}), this isometry takes the form
\begin{equation}
\label{eq:Scale}
    x \rightarrow \xi x \,, \quad \quad y \rightarrow \xi y \,.
\end{equation}
For a spatial region of length $L$ with a radial cutoff $\epsilon$, considering instead a radial cutoff $\xi \epsilon$ is equivalent by isometry to keeping $\epsilon$ fixed and considering the spatial region of length $\xi^{-1} L$; thus going to larger radial distances in the bulk is equivalent to probing shorter distances on the boundary.  In this way, we interpret a radial cutoff in the bulk as an ultraviolet (UV) cutoff in the boundary, providing an upper limit to the wavenumber of possible field excitations. We may think of this as analogous to a lattice regularization, with lattice spacing $a$ directly proportional to $\epsilon$,
\begin{equation} \label{eq:spacing_halfplane}
    a \sim \epsilon \,.
\end{equation}
This correspondence between a bulk radial cutoff and a boundary lattice regularization is not exact, as a lattice also leads to a discrete momentum spectrum, while the UV regularization from a radial cutoff allows for all momenta below the cutoff, and thus terms subleading in the regulator may disagree. However, they share the property that they are local in the sense that the cutoff at a point in the line is independent of any other spatial points that may be participating in the quantity being calculated, such as the other end of the spatial region, or in the case of a correlation function, the locations of other local operators.  This is a natural way to regulate a theory.

\begin{figure}
    \centering
    \begin{subfigure}[m]{0.5\textwidth}
        \centering
        \includegraphics[width=\textwidth]{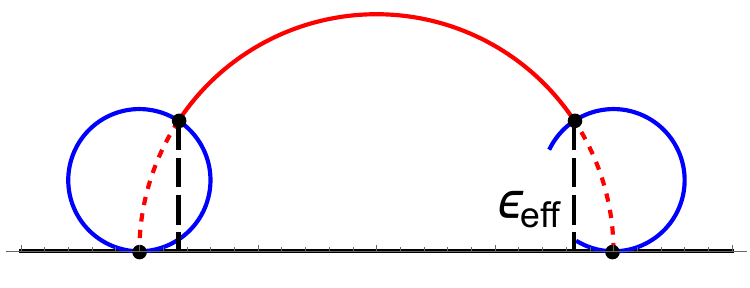}
    \end{subfigure}
    \hfill
    \begin{subfigure}[m]{0.4\textwidth}
        \centering
        \includegraphics[width=\textwidth]{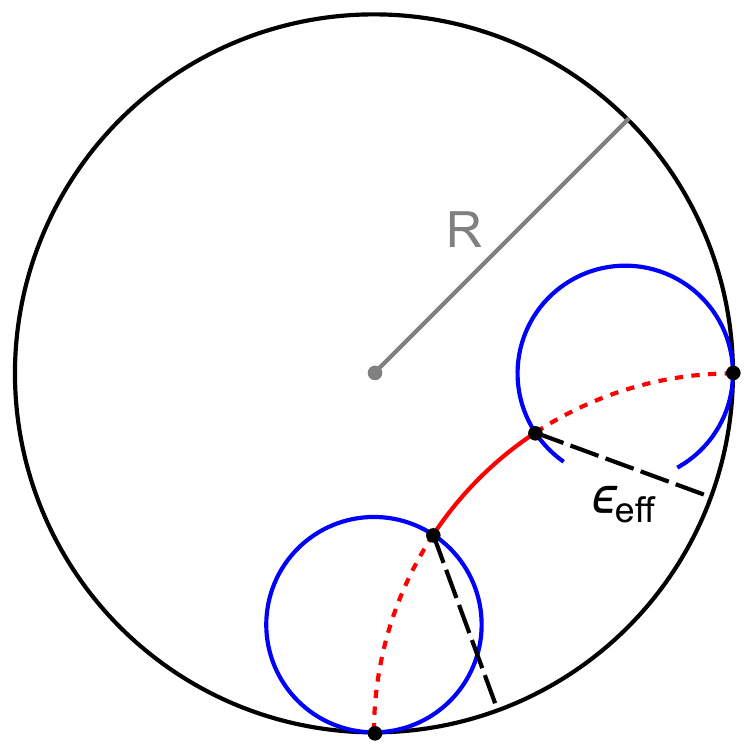}
    \end{subfigure}
    \caption{The effective radial cutoff $\epsilon_\text{eff}$ for the horocycle regularization scheme in the Poincar\'e half-plane (left) and Poincar\'e disk (right). The RT surface corresponding to a boundary subregion is drawn in red, and the horocycle regulators are drawn in blue. $\epsilon_\text{eff}$ is labeled by a black long-dashed line.}
    \label{fig:eps_eff}
\end{figure}

The horocycle scheme, however,  is different.  Consider the entropy of a single subregion of width $L$ on the boundary of a spatial slice of AdS$_3$; as discussed in previous sections, the radial distance to the boundary at which a horocycle terminates an RT surface defines an effective radial cutoff $\epsilon_\text{eff}$ (\ref{HorocycleEffectiveCutoff}) that depends on the location of the far end of the interval.  Viewed as an effective lattice spacing, we have
\begin{equation}
    a \sim \epsilon_\text{eff, plane} = \frac{\epsilon}{1 + (\epsilon/L)^2} \,.
\end{equation}
Because of this $L$-dependence, the boundary dual of horocycle regularization cannot be local -- regularization at one endpoint of the subregion is dependent on how far away the other endpoint is.  For example, decreasing the subregion size $L$ will move the effective radial cutoff closer to the boundary at both endpoints, moving the corresponding lattice cutoff to smaller distances. The horocycle regulator for the upper-half-plane is  illustrated on the left-hand-side of figure~\ref{fig:eps_eff}. 

In the limit where the subregion is large compared to the horocycle ($L\gg\epsilon$) the $L$-dependence disappears, reproducing (\ref{eq:spacing_halfplane}). In general, the constant lattice spacing corresponding to a uniform radial cutoff provides an upper bound for the varying lattice spacing corresponding to a horocycle cutoff. As $L \to \epsilon$ we have $\epsilon_{\rm eff} \to \epsilon/2$, so there is a lower bound as well; of course, we should stop using a cutoff size $\epsilon$ long before $L$ approaches the same value. 

While the horocycle cutoff is very natural from the bulk point of view, its non-local character makes it much more unusual from the field theory point of view. Since the cutoff is defined only for curves, it is not clear whether it should be thought of inducing any reduction in degrees of freedom in the middle of the spatial region; it is only necessary that it cut off the degrees of freedom at the region boundary, where the divergence in the entanglement entropy is. How many modes are cut off at each end point then depends on the distances of the endpoints from each other.

Analogous statements can be made about the Poincar\'e disk model introduced in section~\ref{sec:disk} with metric given by (\ref{eq:disk_metric}); in this geometry,  the lattice spacing is related to the radial cutoff by \cite{ryu_holographic_2006},
\begin{equation} \label{eq:spacing_disk}
    a_\text{disk} \sim \frac{R\epsilon}{2R - \epsilon} \,,
\end{equation}
while the effective radial cutoff scales with $L$ as given in (\ref{eq:effUV_disk}), leading to an effective lattice spacing, 
\begin{equation}
    a_\text{disk} \sim R \frac{1 - \beta}{1 + \beta}, \qquad \beta \equiv \left( 1 + \frac{4\epsilon(\epsilon - 2R)}{4R^2 + \epsilon^2\cot^2(L/2R)} \right)^{1/2} \,.
\end{equation}
Again, in the limit of small cutoff $\epsilon \ll L, R$ the lattice spacing approaches the uniform cutoff result (\ref{eq:spacing_disk}). This case is illustrated on the right-hand-side of figure~\ref{fig:eps_eff}.

\section{Conclusion} \label{sec:conc}

The regularization of holographic entropies is necessary to study the universal properties of these formally infinite quantities. \textit{Finite} information measures (such as the multipartite mutual informations and optimized correlation measures of section \ref{sec:finite}) involve combinations of entropies that lead to cancellations of these divergences, permitting the study of truly finite quantities. In spatial slices of AdS$_3$/CFT$_2$, a unique regulator constructed from horocycles can be used to give expressions for these finite information measures that does not depend on the value of the cutoff, even at finite values. This \textit{totally cutoff-independent} regularization scheme agrees with other well-behaved regulators (such as the uniform radial cutoff) in the limit that the regulator is removed. 

In this paper, we have sought to better understand the properties of the horocycle cutoff scheme, applying it to RT surfaces in both the Poincar\'e half-plane and Poincar\'e disk. We classified a number of finite information measures and demonstrated that all are totally cutoff independent under horocycle regularization. Furthermore, we discussed the non-local properties of possible dual CFT interpretations of horocycle cutoffs, including a boundary lattice with variable spacing depending on the size of the subregion.

While the horocycle regulator is very natural from the bulk point of view, it is non-local on the boundary, cutting off the degrees of freedom at the interface between regions at a scale dependent on the size of the regions. It would be interesting to understand better the proper field theory interpretation of this non-local cutoff.

Additionally, the horocycle regulator provides a natural bulk regularization scheme for geodesic curves, which is the form minimal area surfaces take in two spatial dimensions. It is natural to wonder how the horocycle regulator might generalize to higher dimensions. There are straightforward higher-dimensional generalizations of horocycles called \textit{horospheres}, which are surfaces orthogonal to all geodesic curves approaching a fixed point on the boundary, and thus provide a natural regularization scheme for one-dimensional curves in any spatial dimension. However, in higher dimensions the RT surface calculating an entanglement entropy is a higher-dimensional surface, for which a different geometric regulator would be required. Moreover, in higher dimensions the boundary of the boundary region is a curve or higher-dimensional surface rather than a collection of points, leading to a broader class of possible geometries for the RT surfaces, and thus complicating the question of whether a single simple generalization of the horocycle regulator exists. It would also be interesting to consider horocycle regulators in non-vacuum three dimensional spacetimes such as black holes. Furthermore, horocycles appeared when Hubeny employed bit threads \cite{Freedman:2016zud} to develop an alternate proof of the monogamy of mutual information \cite{Hubeny:2018bri}, and it would be interesting to further explore the connection with the horocycles employed here. We leave these interesting questions to future work.

\section*{Acknowledgements}

We are grateful to Veronika Hubeny and Graeme Smith for helpful discussions. OD and KH are supported by the Department of Energy under grant DE-SC0010005. SA, OD, and KH are supported by the Department of Energy under grant DE-SC0020360.

\bibliographystyle{utphys}
\bibliography{horocycles}

\end{document}